\documentclass[
 aip,
 sd,
 reprint,
 floatfix,
 amsmath,
 amssymb
]{revtex4-2}

\usepackage[utf8]{inputenc}
\usepackage[T1]{fontenc}
\usepackage{graphicx}
\usepackage{url}
\begin{document}

\title{Finite-temperature Fe K-edge X-ray absorption simulations reveal local structural dynamics of an iron(II) photosensitizer in solution and the crystalline phase}

\author{Patrick M\"uller}
\affiliation{Department of Chemistry and Center for Sustainable Systems Design, Paderborn University, Warburger Str.\ 100, 33098 Paderborn, Germany}

\author{Lorena Fritsch}
\affiliation{Department of Chemistry and Center for Sustainable Systems Design, Paderborn University, Warburger Str.\ 100, 33098 Paderborn, Germany}

\author{Matthias Bauer}
\email{matthias.bauer@upb.de}
\affiliation{Department of Chemistry and Center for Sustainable Systems Design, Paderborn University, Warburger Str.\ 100, 33098 Paderborn, Germany}

\author{Thomas D. K\"uhne}
\email{t.kuehne@hzdr.de}
\affiliation{Center for Advanced Systems Understanding (CASUS), Conrad-Schiedt-Strasse 20, 02826 Goerlitz, Germany}
\affiliation{Helmholtz Zentrum Dresden-Rossendorf, Bautzner Landstrasse 400, 01328 Dresden, Germany}
\affiliation{Institute of Artificial Intelligence, Technische Universitaet Dresden, Helmholtzstrasse 10, 01069 Dresden, Germany}

\date{\today}

\begin{abstract}
Interpreting metal K-edge spectra of flexible photosensitizers requires a
structural model that separates electronic signatures from thermal motion,
solvent disorder, and crystal-packing effects. We combine Fe K-edge X-ray
absorption measurements with second-generation Car--Parrinello ab initio
molecular dynamics and all-electron Gaussian and augmented-plane-wave
simulations for an iron(II) N-heterocyclic carbene photosensitizer in
acetonitrile solution and in the crystalline phase. Ensemble-averaged spectra
reproduce the main near-edge features in both environments and preserve the
experimentally observed similarity of the first Fe coordination shell upon
dissolution. Comparison with radial distributions extracted from extended
fine-structure measurements validates the Fe--N and Fe--C coordination shells
sampled by the trajectories, while element-resolved pair distributions explain
why higher-shell experimental contrast is rapidly lost. The same dynamical
ensembles reveal a broad out-of-plane distribution of the terpyridine nitrogen
atom and a nearly octahedral distribution of the Fe-centered coordination
planes. The results show that finite-temperature X-ray absorption simulations
can provide a compact structural-dynamics picture of molecular transition
metal photosensitizers by linking local spectra, solvent-phase ligand motion,
and medium-range structural disorder within one trajectory-based description.
\end{abstract}

\keywords{X-ray absorption spectroscopy, Fe K-edge, HERFD-XAS, EXAFS, ab initio molecular dynamics, structural dynamics, photosensitizers}

\maketitle

\section{Introduction}

X-ray methods provide a direct route to the structural dynamics of molecular
systems because they can probe local electronic structure, coordination
geometry, and transient disorder without requiring long-range periodicity
\cite{CherguiCollet2017,Dau2003,Haumann2005,Glatzel2013}. X-ray absorption
spectroscopy (XAS) is particularly useful for transition metal complexes in
solution, in the solid state, and under working catalytic or photochemical
conditions because the metal K-edge is locally sensitive to the immediate
coordination environment \cite{Bauer2012,Bertagnolli1994}. The near-edge
region, including the pre-edge and X-ray absorption near-edge structure
(XANES), reports on oxidation state, local symmetry, covalency, and
multiple-scattering pathways. The extended X-ray absorption fine structure
(EXAFS) region provides complementary structural information on ligand
identities, coordination numbers, and metal--ligand distances
\cite{Feth2003,Bauer2010}.

The structural information encoded in a transition metal K-edge spectrum is not
only a property of an optimized geometry. For flexible molecular complexes, the
observed spectrum averages over thermal fluctuations, solvent configurations,
and, in the solid state, crystal-packing constraints. These effects are
especially important for photosensitizers, where ligand-field geometry and
metal--ligand covalency control the relative energies of charge-transfer and
metal-centered states. For first-row transition metals, conventional XANES is
further broadened by the 1s core-hole lifetime \cite{Glatzel2005}.
High-energy-resolution fluorescence-detected X-ray absorption spectroscopy
(HERFD-XAS) mitigates this broadening by monitoring a narrow emission window with crystal
spectrometers \cite{Johann1931,Glatzel2005,Bauer2014}. The sharpened
near-edge features increase the structural sensitivity of the experiment but
also make it more important to model finite-temperature and condensed-phase
effects explicitly.

Several electronic-structure strategies are available for core spectroscopy,
including time-dependent density functional theory, many-body perturbation
theory, Bethe--Salpeter treatments, and transition-potential or core-hole
density functional methods
\cite{Shirley1998,Bechstedt2015,Besley2009,DeBeer2008b,Zhang2015,Roper2016}.
For periodic and condensed-phase systems, the all-electron Gaussian and
augmented-plane-wave (GAPW) approach offers a practical route because it
retains an all-electron description near the absorbing atom while preserving a
plane-wave representation of the smooth density between atoms
\cite{Iannuzzi2005,Iannuzzi2007,Iannuzzi2008}. In previous work, combining
GAPW calculations with second-generation Car--Parrinello ab initio molecular
dynamics (AIMD) showed that replacing an isolated gas-phase model by an
explicit crystal or solution environment can change simulated Cu K-edge spectra
as strongly as further changes in the electronic-structure level
\cite{Mueller2019}. It also showed that finite-temperature ensemble sampling is
needed for semi-quantitative spectra of transition metal complexes.

Here we apply this trajectory-based strategy to a photochemically relevant
iron(II) N-heterocyclic carbene (NHC) complex with a
mixed terpyridine--bis(imidazolylidene)pyridine
ligand scaffold that has been investigated as a photosensitizer for
light-driven hydrogen evolution \cite{Zimmer2017}. Specifically, the
complex is [Fe(L)terpy][PF$_6$]$_2$, where terpy denotes
2,2$'$:6$'$,2$''$-terpyridine and L denotes
2,6-bis[3-(2,6-diisopropylphenyl)imidazol-2-ylidene]pyridine
\cite{Zimmer2018IC}. The present focus is not on developing a new core-level
electronic-structure approximation. Instead, we ask whether one finite-
temperature structural ensemble can account for Fe K-edge spectra in solution
and in the crystalline phase and, at the same time, reveal the ligand dynamics
that are hidden behind ensemble-averaged line shapes. This framing is
important for future time-resolved X-ray studies because static structural
models alone cannot distinguish spectral changes caused by electronic
excitation from those caused by thermal and solvent-driven structural disorder.

\section{Methods}

\subsection{Ab initio molecular dynamics}

All calculations were performed with the CP2K program suite
\cite{CP2K2020,Iannuzzi2026}. Atomic configurations were generated at the
density functional theory (DFT) level using the
Becke--Lee--Yang--Parr generalized-gradient approximation \cite{Becke,LYP}.
The mixed Gaussian and plane-wave (GPW) method as implemented in CP2K was used
with separable norm-conserving pseudopotentials \cite{GTH,PP}. Kohn--Sham
orbitals were expanded in a molecularly optimized double-zeta valence plus
polarization (DZVP) Gaussian basis set \cite{MolOpt}, and the plane-wave
cutoff for the electronic density was set to 360~Ry.

The Fe(II) complex was simulated in acetonitrile solution using a periodic cell
containing 473 atoms and in the crystalline phase using a periodic cell
containing 250 atoms. Three-dimensional periodic boundary conditions were used
in both cases. DFT-based AIMD trajectories were generated with the
second-generation Car--Parrinello method as implemented in CP2K
\cite{TDK2007,TDK2014,KuehneProdan2018,HutterIannuzziKuehne2024,CP2K2020,Iannuzzi2026}.
Each system was equilibrated for 25~ps and then propagated for a further
100~ps at 300~K with a time step of 0.5~fs. Trajectories were analyzed with
the TRAVIS program package to obtain radial, angular, and element-resolved pair
distribution functions \cite{Brehm2011}.

\subsection{X-ray absorption simulations}

Fe K-edge X-ray absorption spectra were computed from configurations extracted from the
AIMD trajectories with the all-electron GAPW approach
\cite{Iannuzzi2005,Iannuzzi2007,Iannuzzi2008}. The spectra were calculated
using the full-core-hole (FCH) transition-potential formalism
\cite{Hetenyi,Cavalleri,Prendergast}. In the GAPW representation, localized
Gaussian-type orbitals describe the Kohn--Sham orbitals, while plane waves
represent the smooth density between atoms. The density cutoff was 480~Ry, and
all-electron orbitals were represented with consistent DZVP Gaussian basis sets
developed for solid-state calculations \cite{Peintinger}. The resulting
discrete transitions were convoluted with Gaussian broadening to account for
experimental and lifetime broadening.

\subsection{X-ray absorption measurements}

EXAFS spectra were measured at the XAS beamline of the Angstr\"omquelle
Karlsruhe (ANKA, Karlsruhe, Germany). Measurements were carried out at the
Fe K-edge (7.112~keV) using a Si(111) double-crystal monochromator, and the
energy scale was calibrated with an iron foil. Solid samples were diluted in
cellulose and pressed into pellets, which were measured in transmission mode
in a nitrogen-filled ionization chamber. Solution spectra were collected in
fluorescence mode using a five-element Ge detector and a dedicated liquid
cell. Several spectra were collected and merged to improve the signal-to-noise
ratio.

Fe K-edge HERFD-XAS spectra were collected at beamline ID26 of the European
Synchrotron Radiation Facility (ESRF) in Grenoble, France. Fe K-edge
measurements were performed with a Si(111) double-crystal monochromator at a
ring energy of 6.0~GeV and a ring current between 180 and 200~mA. The measurements
were carried out using two U35 undulators. The incident energy was selected
with the $\langle 111\rangle$ reflection of a double Si-crystal monochromator,
and the energy scale was calibrated with an Fe foil. The incident X-ray beam
had a flux of approximately $2\times10^{13}$ photons~s$^{-1}$ at the sample
position. Solid samples were prepared as wafers with degassed and dried boron
nitride as binder to reduce self-absorption effects. The spectra were recorded
at 20~K in a He cryostat under vacuum conditions. HERFD spectra were measured with an X-ray
emission spectrometer in the horizontal plane. Sample, analyzer crystal, and
avalanche-photodiode detector were arranged in a vertical Rowland geometry.
The Fe K-edge HERFD spectra were obtained by recording the intensity of the
Fe K$\beta_{1,3}$ emission line as a function of the incident energy. The
emission energy was selected using the $\langle 620\rangle$ reflection of five
spherically bent Ge crystal analyzers with a bending radius of 1~m and aligned
at an 80$^\circ$ Bragg angle. Measurements were acquired at several spots, and no
radiation damage was detected within the acquisition time.

\section{Results and discussion}

\subsection{Environment-dependent Fe K-edge spectra}

\begin{figure*}[t]
\centering
\includegraphics[width=0.82\textwidth]{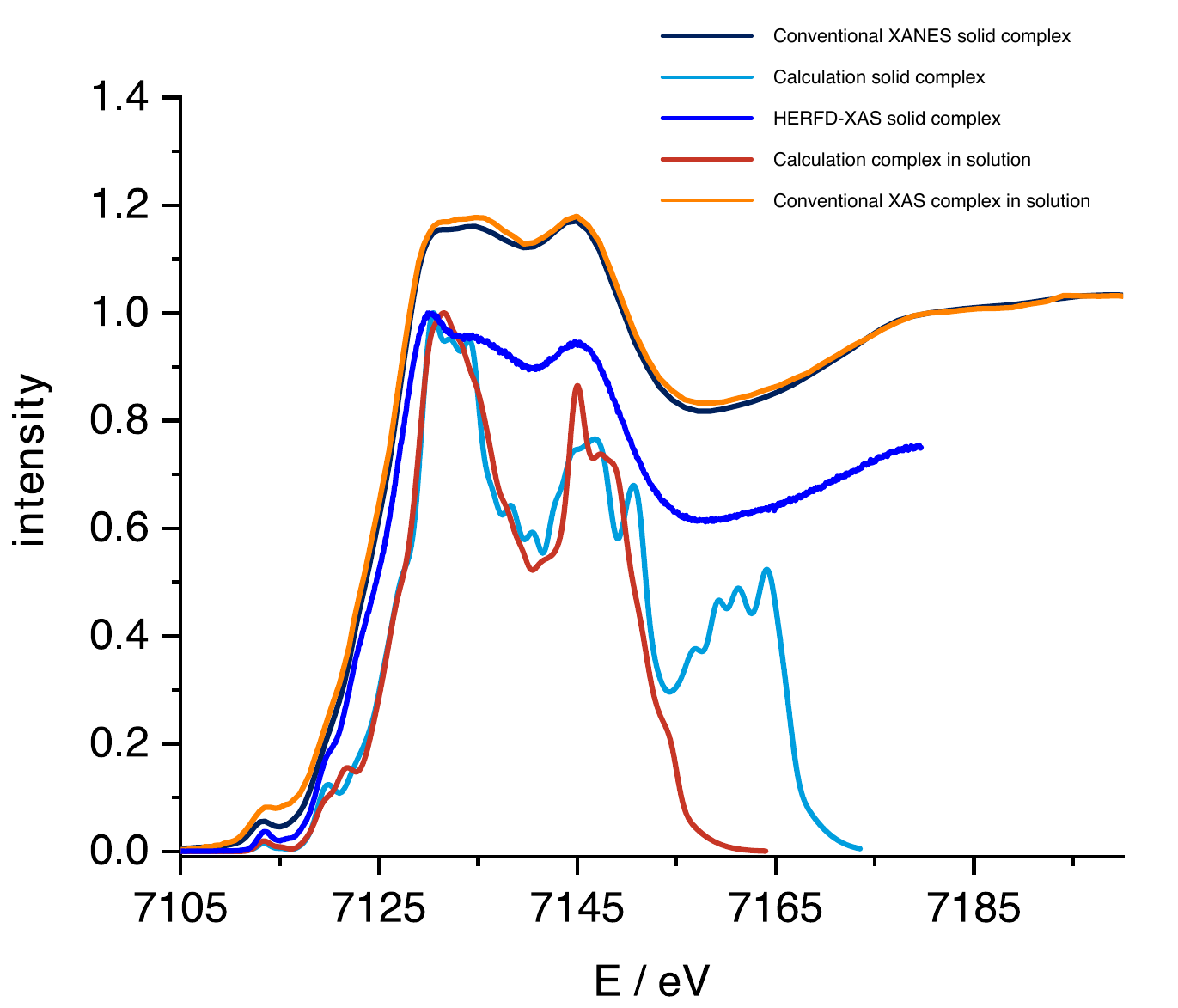}
\caption{Experimental and simulated Fe K-edge XAS spectra of the Fe(II)
complex in acetonitrile solution and in the crystalline phase. Conventional
solid and solution spectra are compared with solid-state
HERFD-XAS and with GAPW/FCH spectra obtained from AIMD
configurations. The conventional solid-state trace corresponds to the
solid-complex XANES spectrum, the high-resolution solid trace to the
HERFD-XAS spectrum, and the conventional solution trace to the complex in
solution; the calculated traces refer to the corresponding solid and solution
AIMD ensembles.}
\label{fig:xas}
\end{figure*}

Figure~\ref{fig:xas} compares experimental and simulated Fe K-edge spectra of
the Fe(II) photosensitizer in acetonitrile solution and in the crystalline
phase. The experimental solid-state spectra recorded in conventional and
high-resolution detection modes show the expected sharpening of near-edge
features in HERFD-XAS. The calculated spectra reproduce the main edge onset
and the dominant near-edge features for both environments. The similarity
between the solution and crystalline spectra indicates that the first Fe
coordination sphere is preserved upon dissolution, while the remaining
differences in the post-edge oscillations reflect changes in medium-range order
and thermal disorder.

The simulations retain sharper oscillatory structure than the experiments,
particularly above the near-edge region. This residual sharpness is expected
for finite ensemble averaging and for a simplified convolution of discrete
transitions. Increasing the energy-dependent broadening would improve the
visual agreement in the post-edge region but would also suppress the
structural sensitivity that makes the near-edge spectrum useful. The relevant
test is therefore not a purely empirical match to the experimental linewidth.
It is whether the same trajectory-based model that describes the edge region
also reproduces independent structural information from EXAFS.

\subsection{Radial structure from experiment and trajectories}

\begin{figure}[htbp]
\centering
\includegraphics[width=\columnwidth]{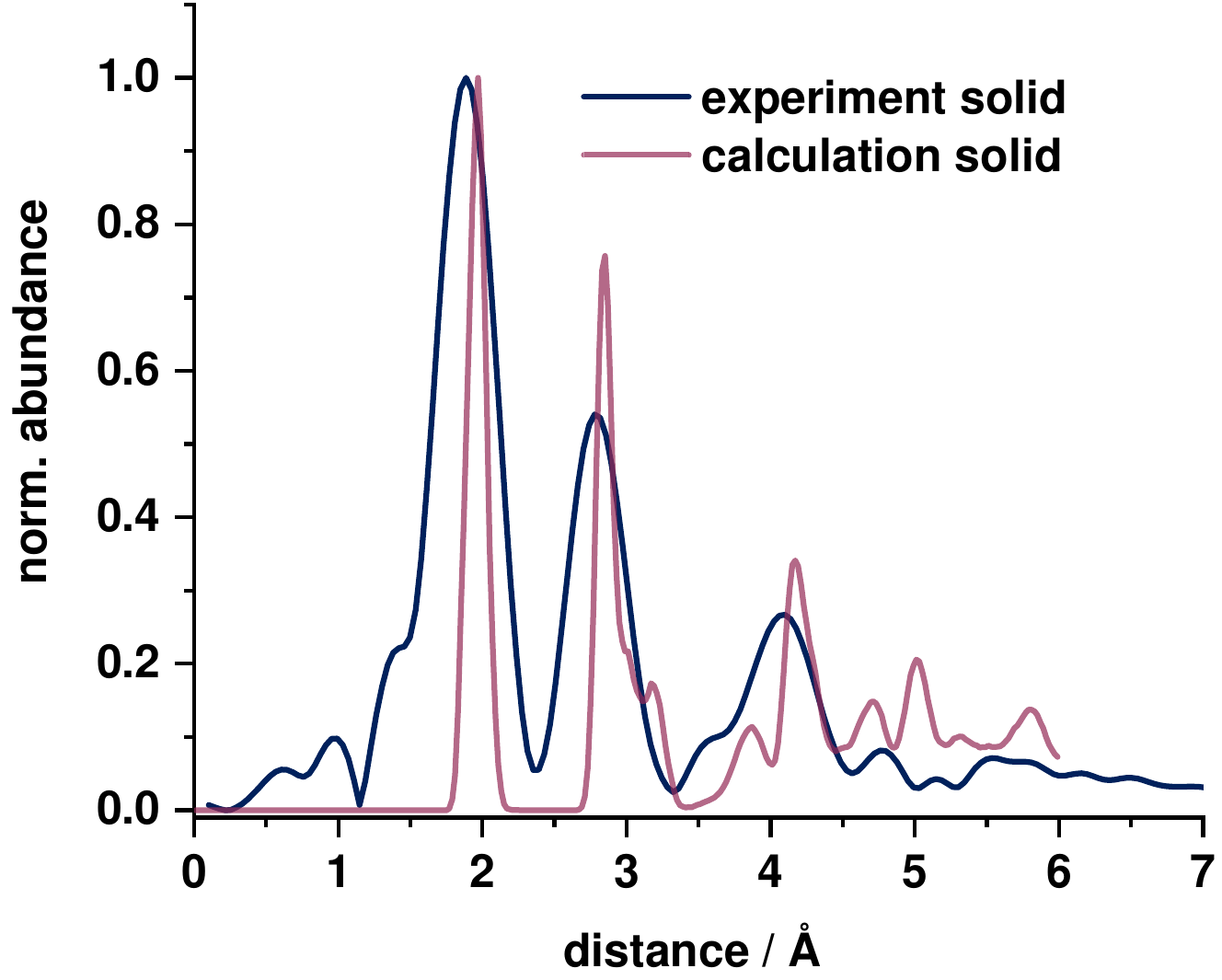}
\caption{Experimental and simulated radial distribution functions of the
Fe(II) complex in the crystalline phase.}
\label{fig:rdf-solid}
\end{figure}

\begin{figure}[htbp]
\centering
\includegraphics[width=\columnwidth]{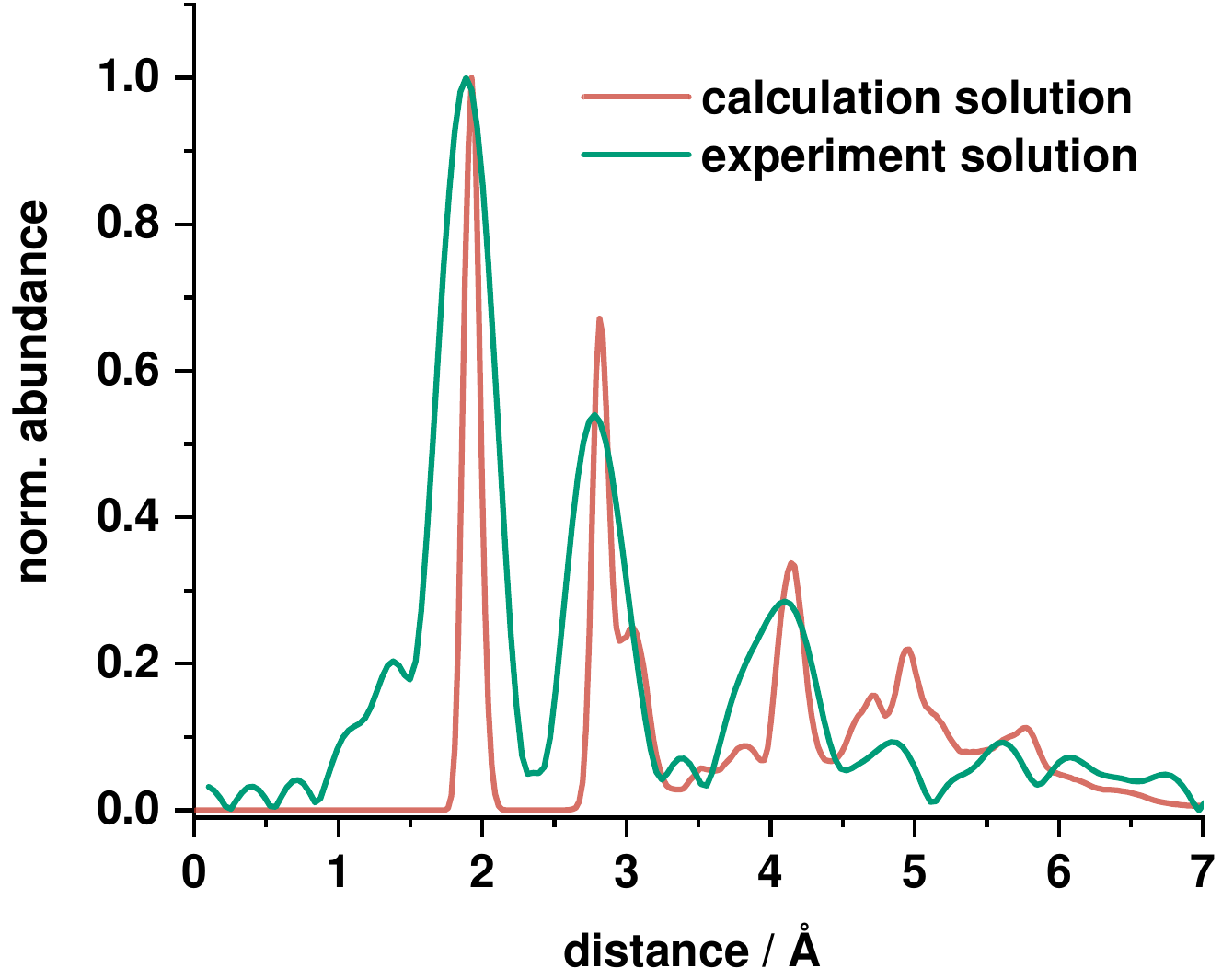}
\caption{Experimental and simulated radial distribution functions of the
Fe(II) complex in acetonitrile solution.}
\label{fig:rdf-solution}
\end{figure}

The EXAFS-derived radial distribution functions (RDFs) provide this independent
validation. Figures~\ref{fig:rdf-solid} and \ref{fig:rdf-solution} compare
experimental and simulated radial distributions in the crystalline and solution
environments. In both cases, the first coordination shell is reproduced close
to 1.9~\AA, and the main higher-shell features around 2.8 and 4.1~\AA\ are
captured by the AIMD trajectories. The calculated distributions are more
structured than the experimental ones because they are obtained directly from
atomistic trajectories, whereas the experimental distributions are broadened by
the finite EXAFS range, disorder, and multiple unresolved scattering
contributions.

\begin{figure}[htbp]
\centering
\includegraphics[width=\columnwidth]{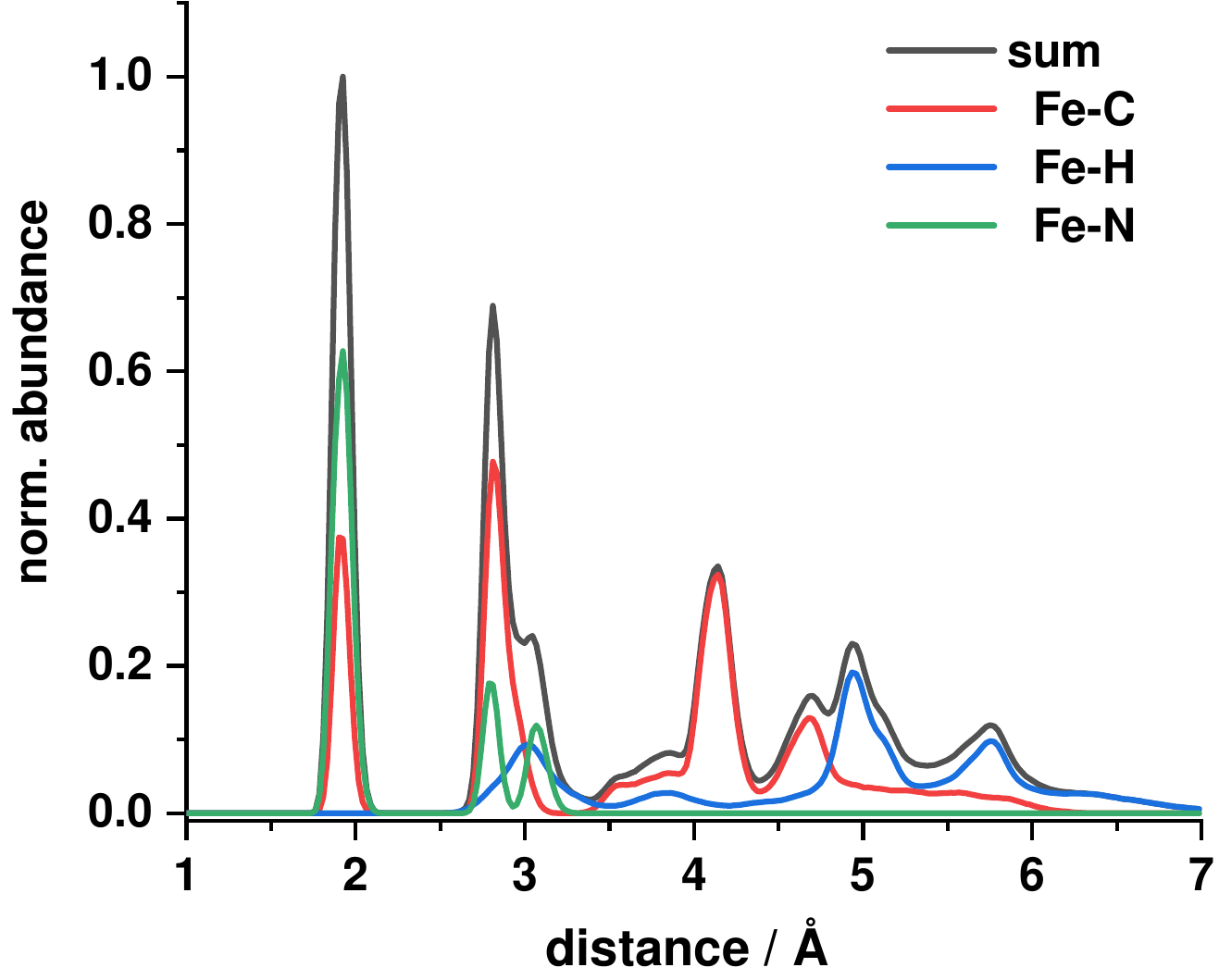}
\caption{Element-resolved simulated radial distribution functions for Fe--C,
Fe--H, and Fe--N correlations in the Fe(II) complex.}
\label{fig:rdf-atom}
\end{figure}

The element-resolved distributions in Fig.~\ref{fig:rdf-atom} show that the
first shell is dominated by Fe--N and Fe--C contributions, consistent with the
NHC and terpyridine coordination environment. At larger distances, the
contribution from light atoms becomes increasingly important. This explains why
the experimental EXAFS-derived structural information fades beyond about
4.5~\AA: the higher-shell signal contains many weak and overlapping
contributions, especially from carbon and hydrogen atoms, which are difficult
to isolate experimentally but are directly accessible in the AIMD analysis.

\subsection{Ligand structural dynamics in solution}

\begin{figure}[htbp]
\centering
\includegraphics[width=\columnwidth]{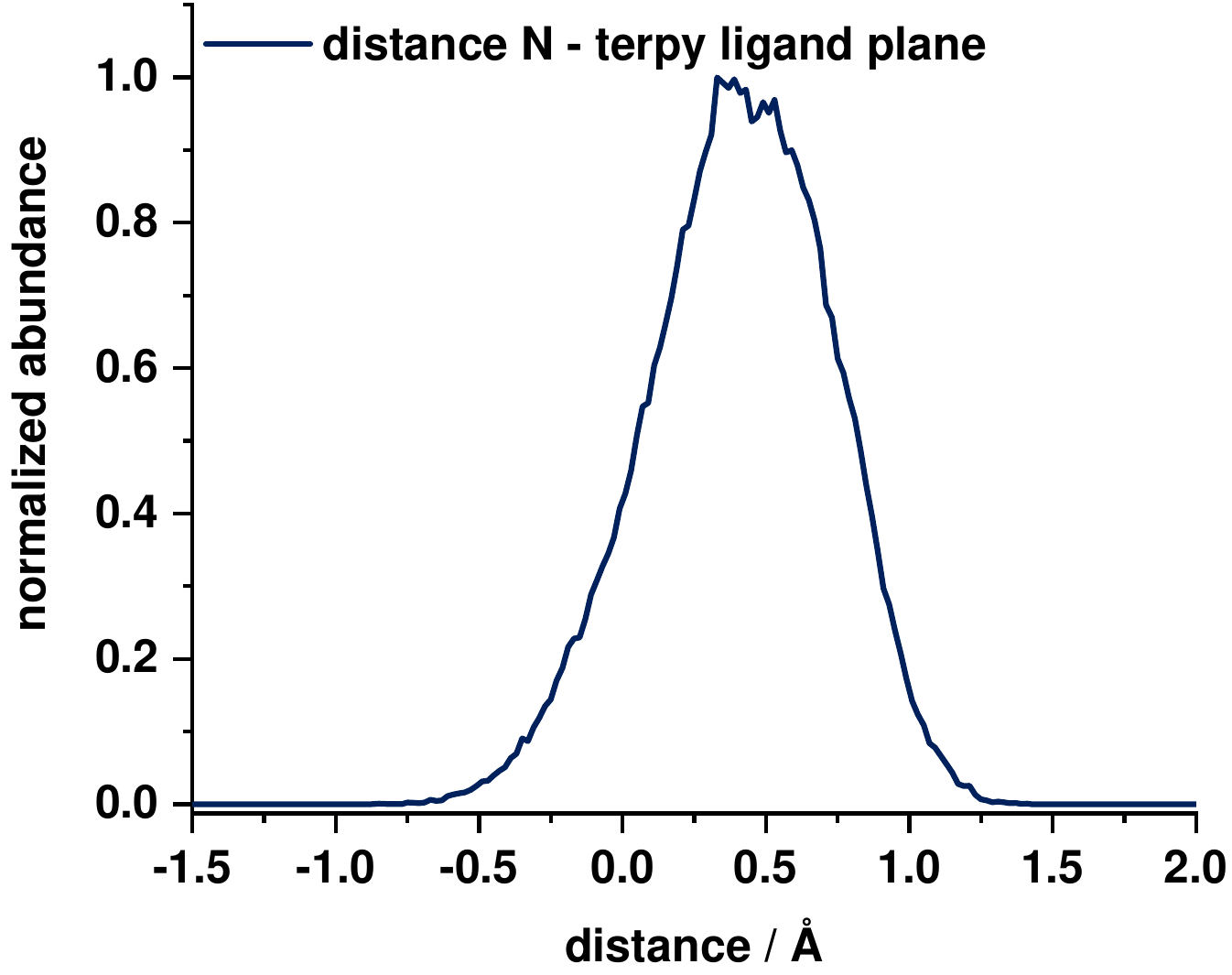}
\caption{Distribution of the distance between the central terpyridine nitrogen
atom and the terpyridine ligand plane in acetonitrile solution.}
\label{fig:terpy}
\end{figure}

Beyond reproducing the local structure that is visible in the Fe K-edge
spectrum, the trajectory analysis provides structural descriptors that are
difficult to extract from a conventional XAS comparison alone. One such
descriptor is the displacement of the central nitrogen atom of the terpyridine
ligand from the terpyridine plane. Figure~\ref{fig:terpy} shows a broad
distribution centered at about 0.4~\AA, with fluctuations extending from
negative displacements to values above 1~\AA. The terpyridine ligand is
therefore dynamically non-planar in solution rather than fluctuating around a
strictly planar reference geometry.

\begin{figure}[htbp]
\centering
\includegraphics[width=\columnwidth]{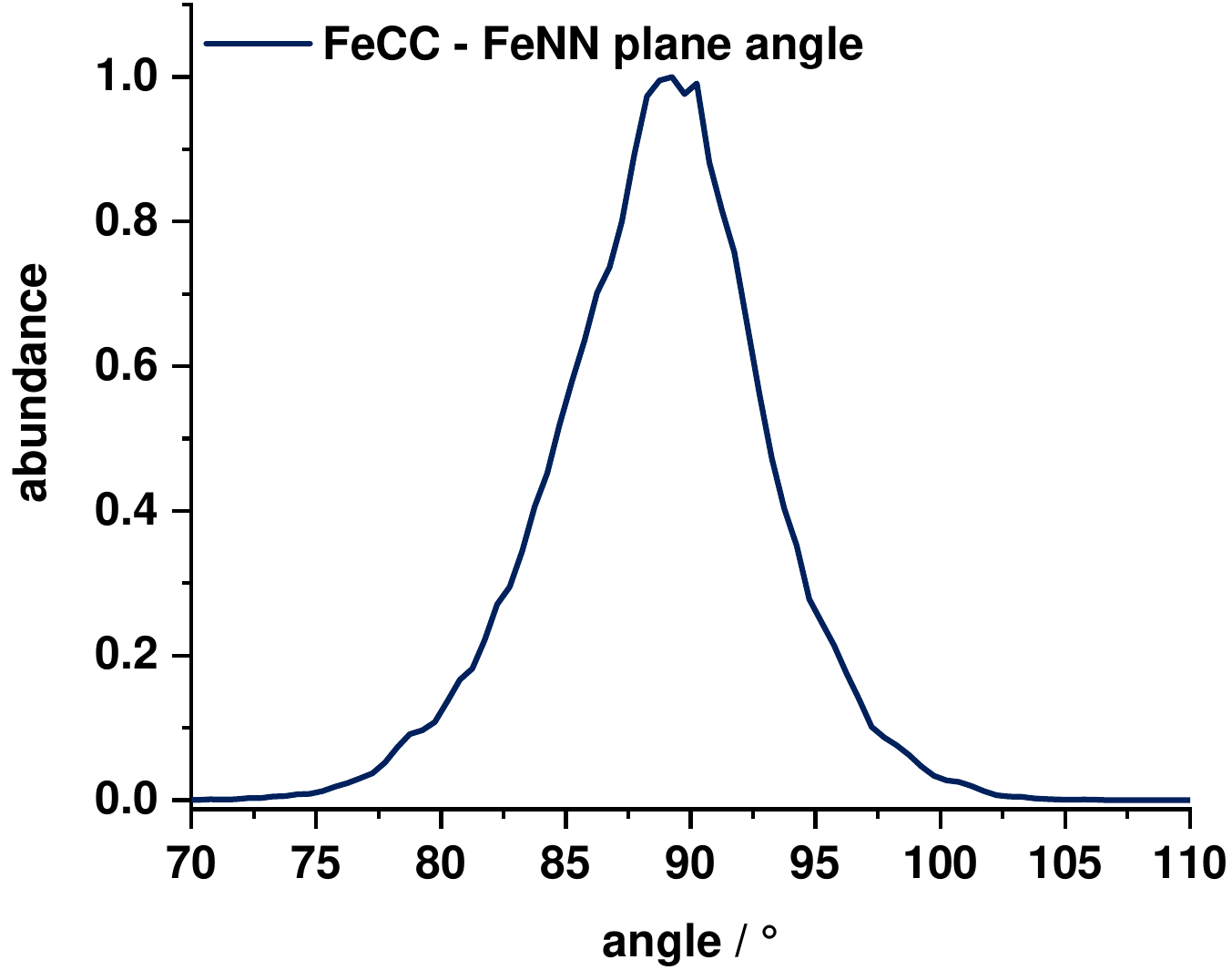}
\caption{Angular distribution between the Fe(NHC) and Fe(terpyridine)
coordination planes in acetonitrile solution.}
\label{fig:angle}
\end{figure}

A second descriptor is the angle between the Fe(NHC) and Fe(terpyridine)
coordination planes. This angle reports on the approximate octahedral geometry
around the iron center, which is relevant for ligand-field splitting and for
the balance between metal-to-ligand charge-transfer and metal-centered states.
As shown in Fig.~\ref{fig:angle}, the distribution is centered close to
89$^\circ$ and remains approximately symmetric around the ideal octahedral
value. Thus, even though the terpyridine ligand itself is flexible, the
Fe-centered coordination environment remains close to octahedral on average.

Taken together, the spectral and trajectory analyses provide a structural-
dynamics interpretation of the Fe K-edge spectra. The near-edge features are
dominated by the preserved first coordination sphere, whereas the radial and
angular distributions expose finite-temperature disorder that is hidden behind
the broadened experimental line shapes. The work therefore extends earlier Cu
K-edge simulations of a relatively rigid, hydrophobic Cu(I) complex
\cite{Mueller2019} to a flexible Fe(II) photosensitizer with a mixed
terpyridine--bis(imidazolylidene)pyridine ligand scaffold in a polar solvent and
shows that the same trajectory can be used both for spectrum generation and for
chemically interpretable structural descriptors.

\section{Conclusions}

We have combined Fe K-edge XAS measurements, second-generation Car--Parrinello
AIMD simulations, and all-electron GAPW/FCH calculations to study an Fe(II)
NHC photosensitizer with a mixed terpyridine--bis(imidazolylidene)pyridine
ligand scaffold in acetonitrile solution and in the crystalline phase. The
computed spectra reproduce the main experimental XANES features and
distinguish the two condensed-phase environments. The remaining
over-structuring of the simulated post-edge region is consistent with finite
configurational sampling and simplified broadening, rather than with a failure
of the local structural model.

Comparison with EXAFS-derived radial distributions validates the local Fe--N
and Fe--C coordination shells sampled by AIMD. Element-resolved pair
distributions explain the loss of experimental EXAFS sensitivity beyond about
4.5~\AA, where many light-atom contributions overlap. The same trajectories
show that the terpyridine ligand is dynamically non-planar in solution, while
the Fe(NHC) and Fe(terpyridine) coordination planes remain centered near an
octahedral 90$^\circ$ arrangement. These results demonstrate that
trajectory-based X-ray absorption simulations can connect Fe K-edge spectral
features with finite-temperature ligand dynamics and local structural disorder
in molecular transition metal photosensitizers.

\clearpage

\begin{acknowledgments}
This project received funding from the European Research Council (ERC) under
the European Union's Horizon 2020 research and innovation programme (grant
agreement No.\ 716142). P.M. and M.B. received funding from FOR 1405 of the
Deutsche Forschungsgemeinschaft. M.B. received funding from the German Federal
Ministry of Education and Research within the project SusChEmX
(FKZ 05K14PP1). We acknowledge sample provision and beamtime at ESRF ID26. The
authors gratefully acknowledge the computing time made available to them on the
high-performance computer Noctua 2 at the NHR Center Paderborn Center for
Parallel Computing (PC2). This center is jointly supported by the Federal
Ministry of Research, Technology and Space and the state governments
participating in the National High-Performance Computing (NHR) joint funding
program (\url{www.nhr-verein.de/en/our-partners}).
\end{acknowledgments}

\section*{AUTHOR DECLARATIONS}

\subsection*{Conflict of Interest}

The authors have no conflicts to disclose.

\subsection*{Author Contributions}

Patrick M\"uller: Investigation, Formal analysis, Visualization, Writing --
original draft. Lorena Fritsch: Investigation, Resources, Writing --
review and editing. Matthias Bauer: Conceptualization, Resources, Supervision,
Validation, Funding acquisition, Writing -- review and editing. Thomas D.
K\"uhne: Conceptualization, Methodology, Software, Supervision, Writing --
review and editing.

\section*{DATA AVAILABILITY}

Raw data were generated at the ANKA and ESRF large-scale facilities. Derived
data supporting the findings of this study are available from the corresponding
authors upon reasonable request.

\end{document}